\documentclass[prl,twocolumn,superscriptaddress]{revtex4-1}

\usepackage[dvipdfmx]{graphicx}
\usepackage{dcolumn}
\usepackage{bm}
\usepackage{color}
\usepackage{amsmath,cases}

\begin{document}
\title{Excitation and relaxation dynamics of spin-waves triggered by\\ultrafast photo-induced demagnetization in a ferrimagnetic insulator}

\author{Yusuke Hashimoto}
\affiliation{Advanced Institute for Materials Research, Tohoku University, Sendai 980-8577, Japan}

\author{Koji Sato}
\affiliation{Institute for Materials Research, Tohoku University, Sendai 980-8577, Japan}

\author{Tom H. Johansen}
\affiliation{Department of Physics, University of Oslo, 0316 Oslo, Norway}
\affiliation{Institute for Superconducting and Electronic Materials, University of Wollongong, Northfields Avenue, Wollongong, NSW 2522, Australia}

\author{Eiji Saitoh}
\affiliation{Advanced Institute for Materials Research, Tohoku University, Sendai 980-8577, Japan}
\affiliation{Institute for Materials Research, Tohoku University, Sendai 980-8577, Japan}
\affiliation{Department of Applied Physics, Faculty of Engineering, University of Tokyo, Tokyo 113-8656, Japan}

\date{\today}

\begin{abstract}

Excitation and propagation dynamics of spin waves in an iron-based garnet film under out-of-plane magnetic field were investigated by time-resolved magneto-optical imaging.
The experimental results and the following data analysis by phase-resolved spin-wave tomography reveal the excitation of spin waves triggered by photo-induced demagnetization (PID) along the sample depth direction.
Moreover, the fast relaxation of PID accompanied by the spin transfer due to spin-wave emission was observed.
Possible scenarios of PID in the garnet film are discussed.
Finally, we develop a model for the spin-wave excitation triggered by PID and explain the magnetic-field dependence in the amplitude of the observed spin waves.

\end{abstract}

\maketitle


Spin waves are the excited states in magnetic materials accompanied by the collective precessional motion of magnetization ($M$).
Nowadays, a number of new ideas for devices using spin waves like logic gates~\cite{Kostylev:2005fua, Schneider:2008fu, Serga2010, Nanayakkara:2014jd}, image analysis ~\cite{Khitun:2010kk}, and reservoir computing for machine learning~\cite{Nakane:2018jc} have been suggested and have attracted great interests due to their practical applications working at room temperature.
In order to realize these devices, the excitation, manipulation, and detection of spin waves are crucial.


Recently, phase-resolved observation of the excitation and propagation dynamics of spin waves in a garnet film has been realized by time-resolved magneto-optical (TRMO) imaging ~\cite{Hashimoto:2017jb,Hashimoto:2018ji,Hashimoto:2018bz, Anonymous:YzAnW49D}.
Spin waves were triggered by illuminating the sample with intense pulsed light.
The use of ultrashort-pulsed light allows instantaneous, non-contact, non-destructive, and non-invasive excitation and observation of spin waves with a sub-pico second time resolution and a micrometer spatial resolution ~\cite{Hashimoto2014}.
The Faraday effect, which rotates the polarization angle of light with an angle proportional to the magnetization along the light propagation direction~\cite{Zvezdin:1997ub}, was used to measure the photo-induced change in $M$ by spin waves.

By analyzing the propagation dynamics of spin waves with a model based on the linear response theory and the Fourier transformation, the dispersion relation of spin waves is reconstructed with their phase information.
This new method is named phase-resolved spin-wave tomography (PSWaT) ~\cite{Hashimoto:2018ji,Hashimoto:2018bz, Anonymous:YzAnW49D}.
Through the detailed investigation by PSWaT, we found that the spin waves were excited through various mechanisms depending on the magnetization configuration ~\cite{Hashimoto:2017jb,Hashimoto:2018ji,Hashimoto:2018bz, Anonymous:YzAnW49D,PhysRevApplied.11.061007}.

In the configuration where our sample has in-plane $M$, spin waves are mainly excited through two processes ~\cite{Hashimoto:2017jb,Hashimoto:2018ji,Hashimoto:2018bz, Anonymous:YzAnW49D}.
One is the magneto-elastic coupling (MEC) caused by the optically-excited elastic waves~\cite{Anonymous:YzAnW49D} while another is attributed to the photo-induced demagnetization (PID)~\cite{Hashimoto:2017jb}.
Spin waves generated by these two processes show very different features in time and space.
The detailed study about spin waves generated by MEC has been reported in the previous studies~\cite{Hashimoto:2017jb, Anonymous:YzAnW49D, Hashimoto:2018ji}.

In this study, we focus on spin waves generated by PID.
The excitation and propagation dynamics of spin waves were observed by TRMO imaging under out-of-plane magnetic fields.
In the experimental data, we directly observe the photo-induced change in $M$ by PID and its recovery, attributed to the spatial propagation of spin-angular momenta due to spin-wave emission.
Finally, we develop a theoretical model about the spin-wave excitation by PID and explain the magnetic-field dependence of the amplitude of the observed spin waves.

\section{Method}

\begin{figure}[h]
\includegraphics[width=8cm]{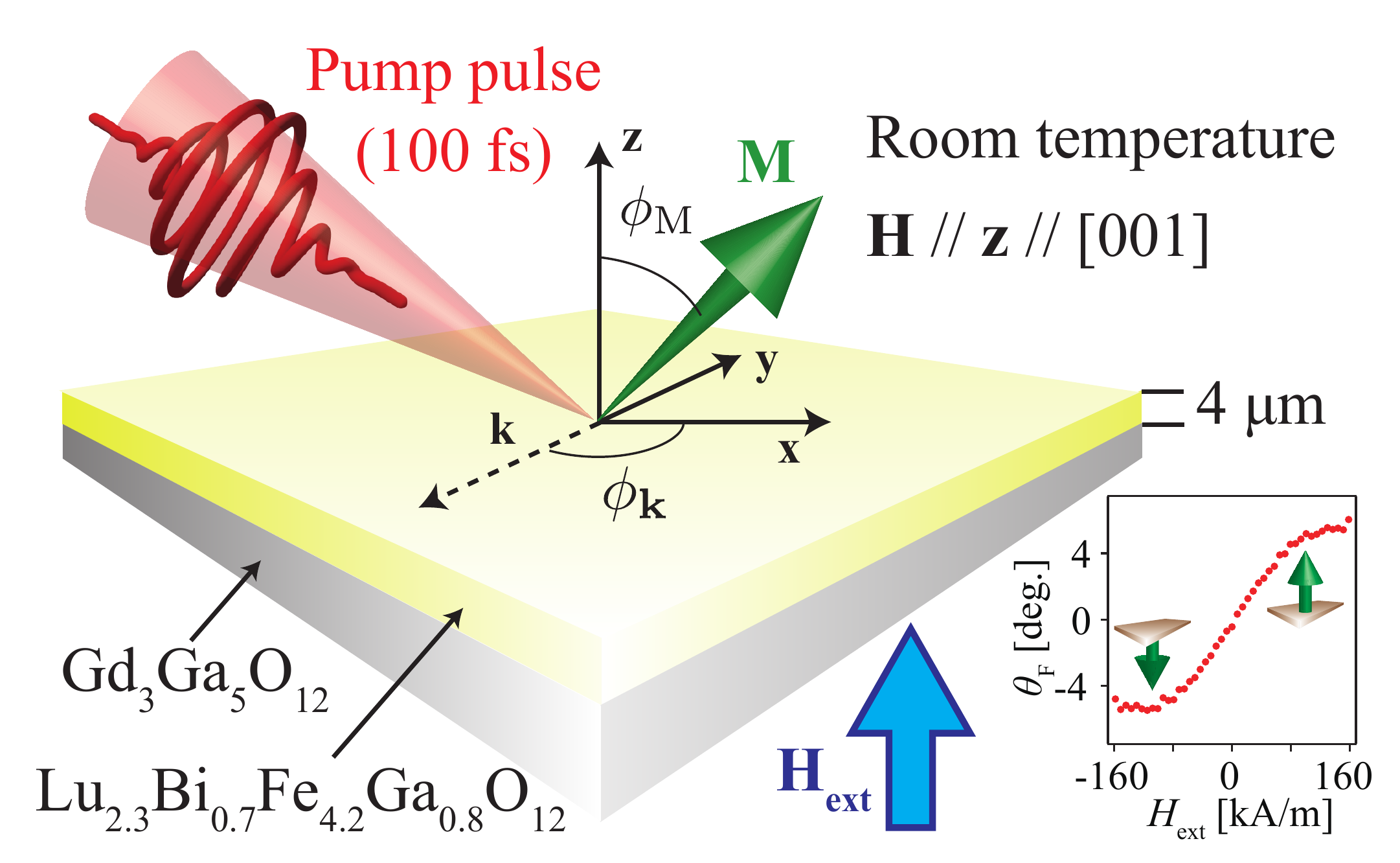}
\caption{\label{FigExeConf}
Schematic illustration of the experimental configuration and the coordinates.
We employ rectangular coordinates ($x$, $y$, $z$). 
The sample surface is in the $x$-$y$ plane, while the surface normal is along the $z$ axis.
The sample has an in-plane magnetization ($M$) in the absence of the external magnetic field ($H_{\rm ext}$), while applying an out-of-plane $H_{\rm ext}$ along the $z$ axis controls the orientation of $M$ in the $x$-$z$ plane.
The angle between $M$ and the $z$ axis is defined as $\phi_{\rm M}$.
The pump pulse is focused on the sample surface at the origin of the coordinate.
The wavevector of the spin wave is denoted as $\bf k$.
The angle between $\bf k$ and the $x$ axis is defined as $\phi_{\bf k}$.
The inset shows the Faraday rotation angle ($\theta_{\rm F}$) as a function of $H_{\rm ext}$.
}
\end{figure}

The excitation, propagation, and relaxation dynamics of optically-excited spin waves were observed with TRMO imaging method.
The experimental configuration and the coordinates ($x$, $y$, $z$) used in this study are schematically drawn in Fig.~\ref{FigExeConf}.
The magnetization vector is denoted by ${\bf M} = (M_{\rm x}, M_{\rm y}, M_{\rm z})$.
The wavevector of the spin wave is denoted as $\bf k$.
The angle between $\bf k$ and the $x$ axis, which is parallel to the in-plane component of $M$, is defined as $\phi_{\bf k}$.

We used a 4-$\mu$m thick Lu$_{2.3}$Bi$_{0.7}$Fe$_{4.2}$Ga$_{0.8}$O$_{12}$ (Bi:LuIG) film grown on a Gd$_{3}$Ga$_{5}$O$_{12}$(001) substrate.
In the absence of the external magnetic field, the Bi:LuIG layer has an in-plane magnetization of 62 kA/m, measured using a vibrating sample magnetometer.
When exposed to a perpendicular applied magnetic field, the vector, $M$, is locally tilted out of plane by an angle that increases with the local field magnitude~\cite{PhysRevB.54.16264}, as is observed in the static Faraday rotation signal ($\theta_{\rm F}$) shown by the circles in the inset of Fig.~\ref{FigExeConf}.
We find that $\theta_{\rm F}$ is proportional to the intensity of the external magnetic field ($H_{\rm ext}$) in the low magnetic field region while is saturated when $H_{\rm ext}$ is larger than 95 kA/m.

As a light source, we used a pulsed laser with 800-nanometer center wavelength, 100-femtosecond time duration, and 1-kilohertz repetition frequency.
The obtained beam was divided to pump and probe beams.
The center wavelength of the probe beam was changed with an optical parametric amplifier to 630 nm, where the sample shows a large Faraday rotation angle (5.2$^{\circ}$) and a high transmissivity (41 $\%$)~\cite{Helseth2001,Hansteen2004}.
The pump beam was circularly polarized and then tightly focused on the sample surface with a radius of approximately 1 $\mu$m.
The pump fluence is estimated to be 480 mJ cm$^{-2}$.
The probe beam was linearly-polarized with a Gran Taylor polarizer and then weakly focused on the sample surface with a radius of roughly 1 mm.
The fluence of the probe beam is almost a hundred times weaker than that of the pump beam.
Optically-excited spin waves were observed through the Faraday effect representing the magnetization component along the sample depth direction [$m_{\rm z}({\bf r}, t)$] by TRMO imaging based on a pump-and-probe technique and a rotating analyzer method using a CCD camera~\cite{Hashimoto2014}.
The spatial resolution of the obtained images is approximately one micrometer, determined by the diffraction limit of the probe beam.
The time delay between the pump and the probe pulses was scanned from -1 ns to 13 ns with the intervals of 0.1 ns.
The detail of the experimental setup is referred to Ref.~\onlinecite{Hashimoto2014}.
All the experiments were performed at room temperature.

\section{Results}

\subsection{Excitation and propagation dynamics of spin waves under out-of-plane magnetic field}

\begin{figure*}
\includegraphics[width=16cm]{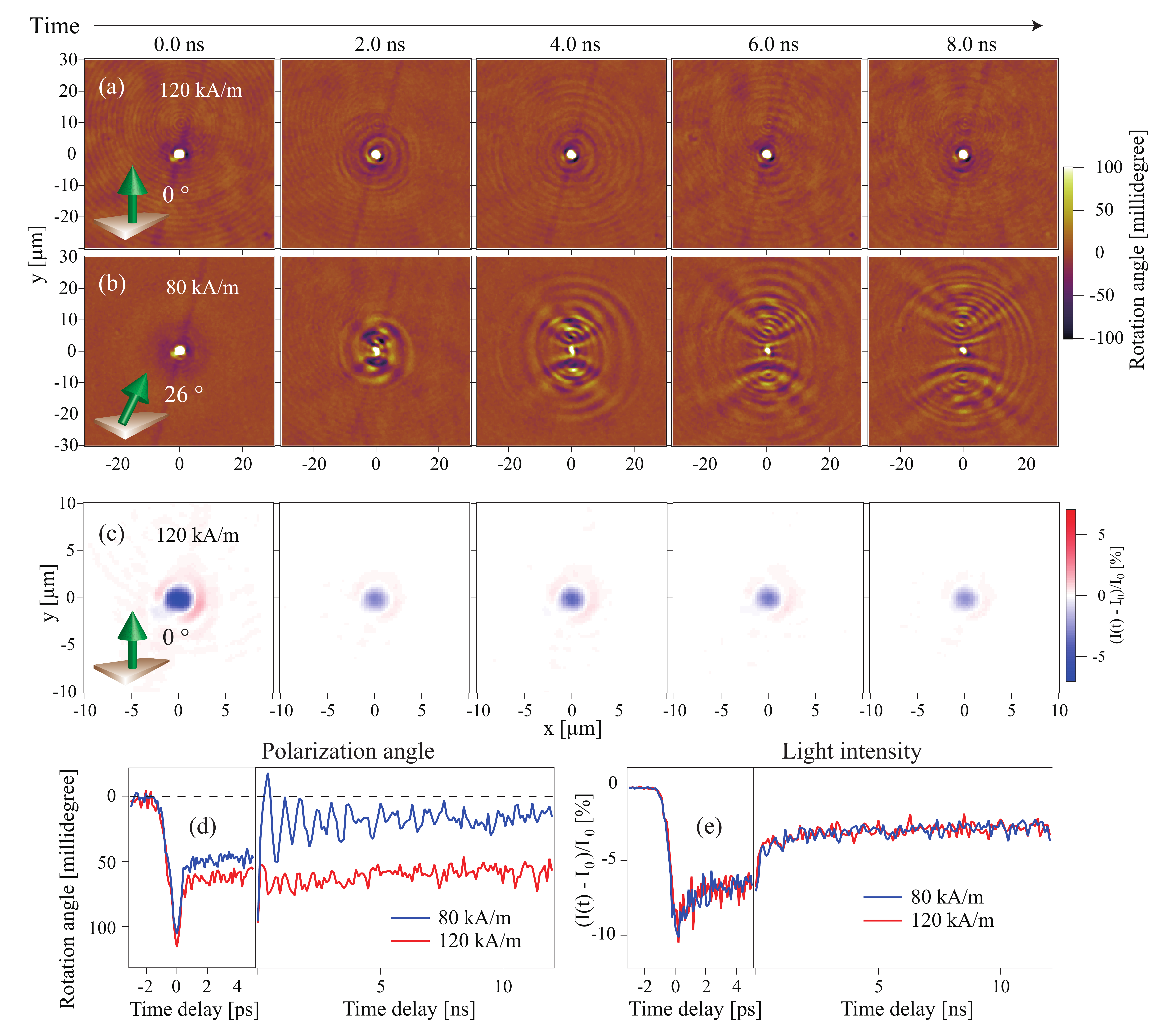}
\caption{\label{FigTRMOILT}
(a, b) TRMO images obtained at the time-delay between pump and probe pulses from 0 ns to 8 ns with the intervals of 2 ns.
The color reflects the photo-induced change in the rotation angle of the polarization plane, proportional to $M_{\rm z}$ through the Faraday effect.
The external magnetic fields of (a) 120 kA/m and (b) 80 kA/m were applied along the sample normal to control the orientation of magnetization as schematically drawn in each figure.
(c) The images of photo-induced change in the intensity of the transmitted light obtained at the time-delay between pump and probe pulses from 0 ns to 8 ns with the interval of 2 ns.
(d, e) The time traces of photo-induced change in the (d) rotation angle of the light polarization and (e) intensity of the transmitted light at the center of the pump pulse excitation.
The red and blue lines are the data obtained under the external magnetic field of 120 kA/m and 80 kA/m, respectively.
}
\end{figure*}

Let us first demonstrate the data obtained with TRMO imaging.
Our system gives the images of the rotation angle of the polarization plane [Figs.~\ref{FigTRMOILT} (a) and (b)] and the intensity [Fig.~\ref{FigTRMOILT} (c)] of the transmitted light, simultaneously.
The image of the rotation angle reflects the sample magnetization through the Faraday effect, while the image of light intensity represents the non-magnetic profile of the sample like the electron distribution and lattice heating.
These interpretations are supported by their different dependences in time, space, and magnetic field.
We also show the temporal change in the signal at the center of the excitation spot in Figs.~\ref{FigTRMOILT}(d) and~\ref{FigTRMOILT}(e).

Figures~\ref{FigTRMOILT}(a) and~\ref{FigTRMOILT}(b) show the images of the polarization angle obtained under the magnetic fields of 120 kA/m and 80 kA/m, respectively.
In these images, we see propagating waves showing strong magnetic field dependence. These are attributed to spin waves through the detailed investigation by PSWaT as shown later.

Spin waves transfer spin angular momenta.
Therefore, the change in $M$, accompanied by the spin-wave emission, is understood by the conservation of the spin-angular momentum.
This is seen in Fig.~\ref{FigTRMOILT}(d).
Under 1.5 kOe, where the spin-wave emission is almost absent, the signal lasts longer than 10 nanoseconds.
On the other hand, under 1.0 kOe where the strong-spin wave were observed, the signal relaxes in several nanoseconds.
This relaxation, showing strong $H_{\rm ext}$ dependence, is attributed to the recovery of $M$ by the spatial propagation of spin-angular momenta due to the spin-wave emission.


The images of light intensity show very different features from those observed in the images of polarization angle.
The signals appear only around the focus of the pump beam and are independent of the external magnetic field, implying their non-magnetic origin.
The temporal change in the signal at the center of the excitation spot [Fig.~\ref{FigTRMOILT}(b)] shows mainly two processes.
One includes exponential decay of sub-nano seconds while another shows signal lasting longer than 10 nanoseconds.
It would be natural to attribute the faster process to the recombination of photoelectrons and photoholes and the slower process to the heating of lattice temperature due to the non-radiative recombination of photoelectrons and photoholes.

\begin{figure}
\includegraphics[width=8cm]{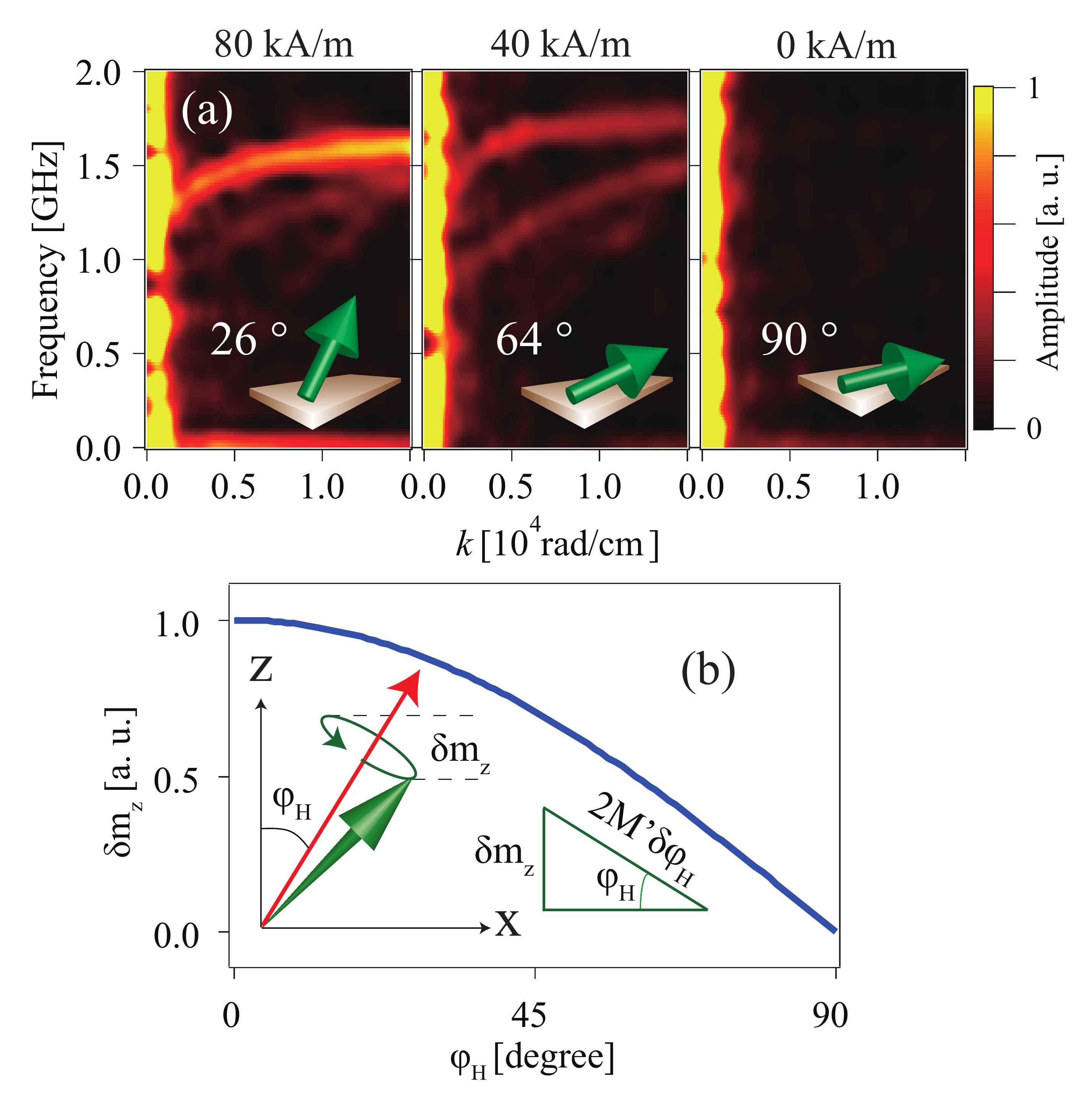}
\caption{\label{FigPSWaTMF}
(a) The magnetic field dependence of the $M_{\rm ee}$ components of the PSWaT spectra along the direction of $\phi_{\bf k} =$ 45 degree.
The magnetic fields of 80 kA/m, 40 kA/m, and 0 kA/m are applied along the sample depth direction.
The angle of $M$ from the surface normal is denoted in each figure.
(b) The relation between $\delta m$ and $\phi_{\rm H}$ calculated with Eq.~\ref{eq:MEWSA5}.
}
\end{figure}

\subsection{Phase-resolved spin-wave tomography}

The observed spin waves were analyzed by PSWaT.
We write the PSWaT spectra, which characterizes spin waves by their spatial symmetry ~\cite{Hashimoto:2018bz}, as $M_{\rm pqr}$; The subscript $p$, $q$, and $r$ reflect the symmetry along $x$, $y$, and time axes, respectively.
Even and odd symmetries are denoted by $e$ and $o$, respectively. 
We also introduce the PSWaT spectra $M_{\rm pq}$, given by $\sqrt{M_{\rm pqe}^2 + M_{\rm pqo}^2}$, to discuss the amplitudes of the spin waves decomposed by their spatial symmetries.
The detailed definition of PSWaT spectra is described in Ref.~\onlinecite{Hashimoto:2018bz}.


Figure~\ref{FigPSWaTMF} shows $M_{\rm ee}$ component of the PSWaT spectra along the $\phi_{\rm k}$ = 45$^{\circ}$ direction.
The external magnetic fields of 0 kA/m, 40 kA/m, and 80 kA/m were applied along the sample depth direction.
The angle of $M$ from the surface normal, estimated from the static Faraday rotation angle, is denoted in each figure.
In the PSWaT spectra, spin-wave dispersions are clearly observed.
The peak frequency is about 1.2 GHz at zero wavenumber ($k$) and increases with increasing $k$.
Above $k \sim 1 \times 10^4$ rad/cm, the increase in the frequency of the peak position saturates.
These are the characteristics of the dispersion relation of spin waves in the low $k$ region, so-called the magnetostatic regime.
We see the spin-wave dispersion in the wide $k$ region.
This fact excludes the magnetoelastic coupling, which resonantly excites spin waves at the crossing of the dispersion relations of spin and elastic waves, from the excitation mechanism of the spin waves.
The amplitudes of spin waves increase with $H_{\rm ext}$.
As shown in Fig.~\ref{FigPSWaTMF}(b), this trend is well explained by the model considering PID, of which detail will be discussed later.

\begin{figure}
\includegraphics[width=9cm]{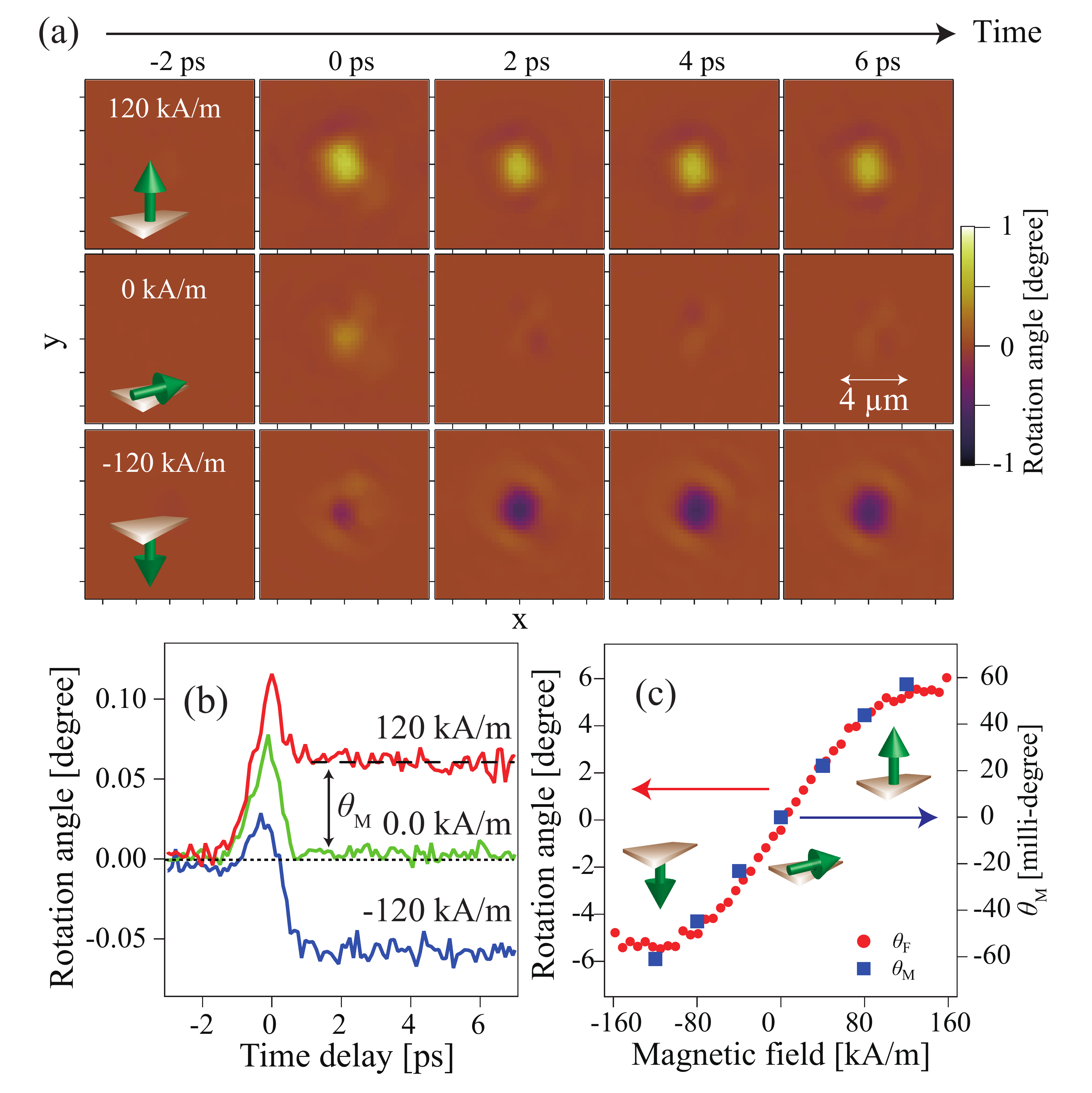}
\caption{\label{FigDemST}
(a) TRMO images obtained at various time delays between pump and probe pulses, shown at top of each figure, and the external magnetic fields of -120 kA/m, 0.0 kA/m, and +120 kA/m applied along the direction normal to the sample surface.
The color represents the photo-induced change in the rotation angle of the polarization plane, reflecting $M_{\rm z}$ through the Faraday effect.
(b) The time traces of the MO signals at the center of the pump pulse excitation.
The red, green, and blue lines are the data obtained under the external magnetic fields of -120 kA/m, 0.0 kA/m, and +120 kA/m, respectively.
To improve the signal-to-noise ratio, the average of the signals in the radius of 0.1 $\mu$m from the center of the excitation spot was calculated.
The dotted line is the fitting result of the data to exclude the magnetic field dependence of $\theta_{\rm M}$, shown in (c).
(c) The magnetic field dependences of $\theta_{\rm F}$ (red circles) and $\theta_{\rm M}$ (blue squares).
}
\end{figure}

In order to reveal the excitation mechanism of the observed spin waves, TRMO images just after the excitation by the pump pulse are shown in Fig.~\ref{FigDemST}(a).
In the images, we see instantaneous change, in the time scale shorter than picosecond, triggered by the pump pulse.
In the time trace shown in Fig.~\ref{FigDemST}(b), we find mainly two components.
One appears only around the time delay of 0 and is independent on $H_{\rm ext}$, implying its non-magnetic origin.
Another, denoted $\theta_{\rm M}$ in the following, appears around 0 and lasts longer than picoseconds.
The intensity of $\theta_M$ shows strong $H_{\rm ext}$ dependence as shown in Fig.~\ref{FigDemST}(c) by the blue squares and is proportional to the static Faraday rotation, reflecting $M_{\rm z}$.
Through the hysteresis-loop experiments of $\theta_{\rm F}$, we confirmed the sign of the signal indicates decrease in $M$.
Therefore, we attribute the change in the polarization angle seen in Fig.~\ref{FigDemST}(a) to the instataneous photo-induced decrease in $M$.
This phenomenan is also referred as ultrafast demagnetization (UFD)~\cite{Beaurepaire:1996es,Hohlfeld:1997cb,Scholl:1997dj,Aeschlimann:1997bx, Regensburger:2000gf,Koopmans:2000dg,Lett:2005ii, Wang:wv, Koopmans:2010ds}.

\subsection{Mechanism of photo-induced demagnetization}

After the first observation of UFD in a Ni film~\cite{Beaurepaire:1996es}, UFD has been observed in various magnetic materials including ferromagnetic metals, such as Ni~\cite{Beaurepaire:1996es,Hohlfeld:1997cb,Scholl:1997dj,Regensburger:2000gf,Koopmans:2000dg,Koopmans:2010ds}, Co~\cite{Aeschlimann:1997bx}, and metallic compounds~\cite{Lett:2005ii}, ferromagnetic semiconductor InMnAs~\cite{Wang:wv}, but not in magnetic insulators.
Extensive efforts have been devoted to reveal the detailed mechanism of UFD theoretically~\cite{Zhang:2000bm,Bigot:2009bu} and by numerical simulations~\cite{Krieger:2015io}  .
However, its understanding has been still under debate.
The difficulty is mainly due to the impulsive nature of UFD, which limits available tools for the investigation.
In the following, we discuss the possible origin of UFD suggested by the data shown in the present study.

In the previous studies, the mechanism of UFD has been discussed by three temperature model~\cite{Beaurepaire:1996es, Koopmans:2010ds}.
The temperature of spin, lattice, and electron systems are defined independently.
The energy transfer between each system is treated as the change in their temperatures.
In this model, PID is treated as the increase of the spin temperature.
Therefore, the key issue is how the spin temperature is heated up by the pump beam.
In the following, we consider the three possible scenarios considering spin-lattice coupling, opto-magnetic effects, and the spin-dependent charge-transfer transition accompanied by spin relaxation of photoholes.

Let us first consider the heating of the spin system through the spin-lattice coupling.
The optical-excitation of the sample creates photoelectrons and photoholes and then heats up the lattice temperature due to their non-radiative recombination.
The energy transfer from lattice to spin system through the spin-orbit coupling results in the heating of the spin system, and thus can be the origin of PID.
However, this is the reverse process of the spin relaxation through the spin-lattice coupling.
The long-living spin waves lasting longer than 10 ns convince us the very weak spin-lattice coupling in the present case of the iron garnets so that this scenario is safely excluded in our case.

Second, we also exclude the direct excitation of the spin system by opto-magnetic effect~\cite{Kimel:2005gy}, which transfers the spin-angular momenta of circularly-polarized light to the spin system.
This is convinced because the sign of the observed spin waves is independent of the helicity of the pump beam.

Finally, we consider the spin-dependent photo-induced charge-transfer transition (CTT) accompanied by the transfer of the spin-angular momenta from spin system to lattice via the spin relaxation of photoholes.
The photo-induced change in $M$ by PID is proportional to the intensity of the pump beam, indicating the contribution of the linear absorption.
In the visible wavelength regime, iron garnets show large absorption attributed to CTT between 2d states of oxide ions and 3d states of iron ions~\cite{Zvezdin:1997ub}.
Although the resonance frequency of this transition ($\sim$3 eV) is roughly twice the photon energy of the pump beam used in this study ($\sim$1.5 eV), this process is not negligible because of the broad absorption spectrum of this transition~\cite{Hansteen2004}.
Since $M$ of the iron garnet film is originated from the 3d states of the iron ion, CTT related to 3d states of iron ions can change $M$.

In this process, the recombination of the photoelectrons and photoholes, which occur in the time scale shorter than nanoseconds in general~\cite{Shah:2013di}, relaxes the induced change in $M$.
Then, the time scale of this process cannot explain the long-living change in $M$ shown in Fig.~\ref{FigTRMOILT}(d).
Therefore, we consider the transfer of the spin-angular momenta from $M$ to lattice through the spin relaxation of photoholes in oxide ions.
If the spin relaxation time of photohole is faster than the recombination time of photoelectrons and photoholes, photoholes recombine with 3d electrons of iron ions with up or down spins.
This breaks the spin orientation of the 3d states of iron ions and thus can be the mechanism of PID.

In this scenario, the spin relaxation of photoholes in oxide ions is crucial.
The spin-relaxation time of photoholes can be shortened by the enhancement of spin-orbit coupling by Bi ions doped to increase the magneto-optical activity.
Therefore, the systematic investigation of PID as a function of the Bi composition is highly expected.
However, the sample preparation of iron-garnet films with different Bi compositions is very difficult.
Moreover, the change in the dopant composition also change the magneto-optical activity and the magnetic properties of the sample.
Therefore, the use of other methods like X-ray Magnetic Circular Dichroism method, which realizes the element specific observation of electron and spin dynamics~\cite{Stamm:2007hy,Boeglin:2010dz}, and first-principle calculation ~\cite{Krieger:2017ho} will be very important as the case of the studies about UFD in ferromagnetic metals.

\subsection{Modeling}


As shown in Fig.~\ref{FigPSWaTMF}(a), the amplitudes of the observed spin waves strongly depend on the external magnetic field, applied to change the orientation of $M$.
The observed trend is well explained, as shown in Fig.~\ref{FigPSWaTMF}(b), with a model considering PID shown below.

Let us consider the free-energy density of the magnetic system~\cite{Dreher2012} given by
\begin{eqnarray}
\label{eq:MEWS5}
G = -\mu_{\rm 0} {\bf H}_{\rm ext} \cdot {\bf m} + B_{\rm d} m_{\rm z}^2 + B_{\rm u}({\bf m} \cdot {\bf u})^2 - \mu_{\rm 0}{\bf H}_{\rm ex} \cdot {\bf m}, \nonumber\\ 
\end{eqnarray}
where ${\bf m}$ is the unit vector ${\bf m}$ = ${\bf M}$/$M_{\rm s}$ = ($m_{\rm x}$, $m_{\rm y}$, $m_{\rm z}$), where $M_{\rm s}$ is the saturation magnetization.
The terms in the right-hand side reflect Zeeman coupling, shape anisotropy ($B_{\rm d}$), magnetocrystalline anisotropy ($B_{\rm u}$), and exchange coupling (${\bf H}_{\rm ex}$), respectively.
We applied a magnetic field along the sample depth direction so that ${\bf H}_{\rm ext} = H_{\rm ext}{\bf z}$.
Since the present sample has an uniaxial magnetic anisotropy ($K_{\rm u}/M_s$ = -39 mT) much larger than the cubic magnetic anisotropy ($K_{\rm c}/M_s$ = 7.5 mT), we disregard $K_{\rm c}$ for simplicity.
For a single domain structure, the exchange coupling is negligible.
Then, Eq.~(\ref{eq:MEWS5}) is simplified to
\begin{eqnarray}
\label{eq:MEWS6}
G = -\mu_{\rm 0} H_{\rm ext} m_{\rm z} + (B_{\rm d} + B_{\rm u}) m_{\rm z}^2.
\end{eqnarray}
Defining the angle between the magnetization and the $z$-axis as $\phi_{\rm M}$ ($m_z = \cos\phi_{\rm M}$), $B_{\rm d}$ = $\mu_{\rm 0}M_{\rm s}/2$, and $B_ {\rm u}$ = $\mu_{\rm 0}H_{\rm u}/2$, we write 
\begin{eqnarray}
\label{eq:MEWS10}
G &=& -\mu_{\rm 0} H_{\rm ext} \cos\phi_{\rm M} + (B_{\rm d} + B_{\rm u}) \cos^2\phi_{\rm M}.
\end{eqnarray}
Then, the angle of magnetization at equilibrium ($\partial G/\partial \phi_{\rm M} = 0$) is 


\begin{subnumcases}
{\label{eq:MEWS14}
\phi_{\rm M} =}
0 & $\frac{H_{\rm ext}}{M_{\rm s} + H_{\rm u}} > 1$\\
\cos^{-1} \left( \frac{H_{\rm ext}}{M_{\rm s} + H_{\rm u}} \right) & $|\frac{H_{\rm ext}}{M_{\rm s} + H_{\rm u}}| \leq 1$\\
\pi & $\frac{H_{\rm ext}}{M_{\rm s} + H_{\rm u}} < -1$.
\end{subnumcases}
The $H_{\rm ext}$ dependence of $\theta_{\rm F}$ [$\propto \cos\phi_{\rm M}(H_{\rm ext})$] is well explained by Eqs.~(\ref{eq:MEWS14}a)-(\ref{eq:MEWS14}c) with the parameters obtained in Ref.~\cite{Hashimoto:2017jb} and the saturation Faraday rotation angle of $5.1^{\circ}$.

The optical excitation of spin waves has been discussed in terms of the photo-induced change in the effective magnetic field ${\bf H}_{\rm eff}$.
We denote the angle between ${\bf H}_{\rm eff}$ and the $z$ axis as $\phi_{\rm H}$.
The direction between ${\bf M}$ and ${\bf H}_{\rm eff}$ is parallel at equilibrium ($\phi_{\rm M}$ = $\phi_{\rm H}$), while the photo-induced change in $|\bf M|$ can change $\phi_{\rm H}$.
Then, magnetization feels torque, $-\gamma {\bf M} \times {\bf H}_{\rm eff}$, and starts to precess about ${\bf H}_{\rm eff}$.
$\phi_{\rm H}$ after the pump pulse illumination may be calculated by replacing $M_{\rm s}$ by $M'({\bf r}, t) (= |{\bf M}({\bf r}, t)|$ = $M_{\rm s}$ + $\delta M({\bf r}, t))$ as 
\begin{eqnarray}
\label{eq:MEWSA3}
\phi_{\rm H}({\bf r}, t) = \cos^{-1} \left( \frac{H_{\rm ext}}{M_{\rm s} + \delta M({\bf r}, t) + H_{\rm u}}\right).
\end{eqnarray}
This model is justified in the duration much shorter than the damping of the precessional motion of the magnetization.
Here, we assume the dominant contribution in the denominator component of Eq.~(\ref{eq:MEWSA3}) by the photo-induced change in $M_{\rm s}$ since $M_{\rm s}$ (78 mT) is twice as great as the intensity of $H_{\rm u}$ (-39 mT).
The exchange coupling caused by the gradient of magnetization by PID is disregarded because of the large radius of the pump focus ($\sim1 \mu m$) and the small change in the magnetization ($< 10 \%$), resulting in the small gradient of magnetization.
Then, the photo-induced change in $\phi_{\rm H}({\bf r}, t)$ is given by
\begin{eqnarray}
\label{eq:MEWSA4}
\delta \phi_{\rm H}({\bf r}, t) &=& \frac{\partial \phi_{\rm M}({\bf r}, t)}{\partial M} \delta M({\bf r}, t)\\
&=& \frac{\cot\phi_{\rm M}({\bf r}, t)}{M_{\rm s} + H_{\rm u}} \delta M({\bf r}, t).
\end{eqnarray}
This equation is applicable when $\delta \phi_{\rm H} \ll \phi_{\rm M}$.
In our experiments, spin waves are observed through the Faraday effect, reflecting $M_{\rm z}$, so that the observed precession amplitude of the magnetization in the spin wave ($\delta m_{\rm z}$) for $0^\circ < \phi_{\rm M} \leq 90^{\circ}$ is
\begin{eqnarray}
\label{eq:MEWSA5}
\delta m_{\rm z} \approx 2M'\delta \phi_{\rm H}\sin\phi_{\rm M} \propto \cos\phi_{\rm M},
\end{eqnarray}
which is plotted in Fig.~\ref{FigPSWaTMF}(b).
The result indicates that the amplitude of spin waves is zero for $\phi_{\rm M} = 90^{\circ}$, and it increases as $\phi_{\rm M}$ decreases from 90$^{\circ}$.
These features show good agreement with the experimental data shown in Fig.~\ref{FigPSWaTMF}(a).
When $H_{\rm ext} =$ 120 kA/m is applied (data not shown), spin waves are not excited because the sample magnetization is fixed along the surface normal since $|H_{\rm ext}/(M_{\rm s} + H_{\rm u})|$ is larger than unity [see Eq.~(\ref{eq:MEWS14})].
For a field strong enough to satisfy the condition, the direction of the magnetization is always along the $z$ direction even after the saturation magnetization changes, and consequently there will be no torque acting on the magnetization.


\section{Summary}

The excitation and propagation dynamics of spin waves generated by photo-induced demagnetization (PID) in an iron garnet film was investigated by time-resolved magneto-optical imaging.
The generation of spin waves by PID was convinced through the systematic experiments under the out-of-plane external magnetic field.
We observed the fast recovery of PID and attributed it to the spin angular momentum transfer due to the spin-wave emission.
A model for the spin-wave emission by PID was developed and explains the magnetic field dependence of the amplitude of spin waves observed by our experiments.




Rare-earth iron garnet has widely been used in spintronics and magnonics devices due to its very small magnetic damping and long spin relaxation time.
Slow magnetic relaxation is accompanied by slow magnetic response, which limits the operation frequencies of spin-wave devices.
The instantaneous manipulation of magnetization suggested in the preset study could overcome this limitation and may lead to applications working for future fast spintronics and magnonic devices.

\begin{acknowledgments}
We thank Mr. T. Hioki, and Dr. R. Ramos for fruitful discussions.
This work was financially supported by JST-ERATO Grant Number JPMJER1402, and World Premier International Research Center Initiative (WPI), all from MEXT, Japan.
\end{acknowledgments}


%

\end{document}